\documentclass{iau} 
\usepackage{graphicx}

\title[NGC 300 ULX1] 
{NGC 300 ULX1: A new ULX pulsar in NGC 300}

\author[CM, SC, FH, GV]   
{Chandreyee Maitra,$^1$
Stefania Carpano,$^1$
Frank Haberl$^1$
 \and Georgios Vasilopoulos$^1$  \thanks{Present address: Yale Department of Astronomy
P.O. Box 208101, New Haven, CT 06520-8101, USA},}  

\affiliation{$^1$Max-Planck-Institut f\"{u}r extraterrestrische Physik, Giessenbachstra{\ss}e 1, 85748 Garching, Germany }

\pubyear{2019}
\volume{XXX}  
\jname{IAU Symposium 346: High-mass X-ray binaries: 
illuminating the passage from massive binaries 
to merging compact objects}
\editors{L. Oskinova, E. Bozzo \& T. Bulik, eds.}

\newcommand{\xmm}{{\it XMM-Newton}}
\newcommand{\nus}{{\it NuSTAR}}
\newcommand{\swi}{{\it Swift}}

\newcommand{\ngc}{NGC\,300\,ULX1}

\newcommand{\expo}[1]{$\times 10^{#1}$}

\newcommand{\msun}{$M_{\odot}$}
\newcommand{\ergcm}[1]{erg cm$^{-2}$ s$^{-1}$}
\newcommand{\ergs}[1]{erg s$^{-1}$}

\begin{document}

\maketitle

\begin{abstract}
\ngc\ is the fourth to be discovered in the class of the ultra-luminous X-ray pulsars. Pulsations from \ngc\ were discovered
during simultaneous \xmm\ / \nus\ observations in Dec. 2016.  The period decreased from 31.71\,s  to 31.54\,s within a few days, with a spin-up rate 
of $-5.56\times10^{-7}$\,s\,s$^{-1}$,  likely one of the largest ever observed from an accreting neutron star.  Archival \swi\ and {\it NICER} observations revealed that the period decreased exponentially from $\sim$45 s to $\sim$17.5 s over 2.3 years. The pulses are highly modulated with a pulsed fraction strongly increasing with energy and reaching nearly 80\% at energies above 10\,keV.  The X-ray spectrum is described by a power-law and a disk black-body model, leading to a 0.3--30\,keV unabsorbed luminosity of $4.7\times10^{39}$ erg\,s$^{-1}$. The spectrum from an archival \xmm\ observation of 2010 can be explained by the same model, however, with much higher absorption. This suggests, that the intrinsic luminosity did not change much since that epoch. \ngc\ shares many properties with supergiant high mass X-ray binaries, however, at an extreme accretion rate.
\end{abstract}

\firstsection 
\section{Introduction}
Ultra-luminous X-ray sources (ULXs) are point-like non-nuclear sources that emit at luminosities in excess of $\sim$ 10$^{39}$\, \ergs, which is approximately the Eddington limit for a spherically 
symmetric accretion onto a stellar mass black hole of 10\msun. Although initially believed to harbour super-critically accreting stellar mass or intermediate mass black holes in order to support the exceedingly high super-Eddington luminosity, recent years have provided undisputed evidence that a substantial fraction of ULXs harbour highly magnetized accreting neutron stars (\cite[Bachetti \etal\ 2014]{Bachettietal2014}, \cite[F\"{u}rst \etal\ 2016]{furst2016}, \cite[Israel \etal\ 2017]{Israel2017}). 

\begin{figure}[b]
\begin{center}
 \includegraphics[width=5.6in]{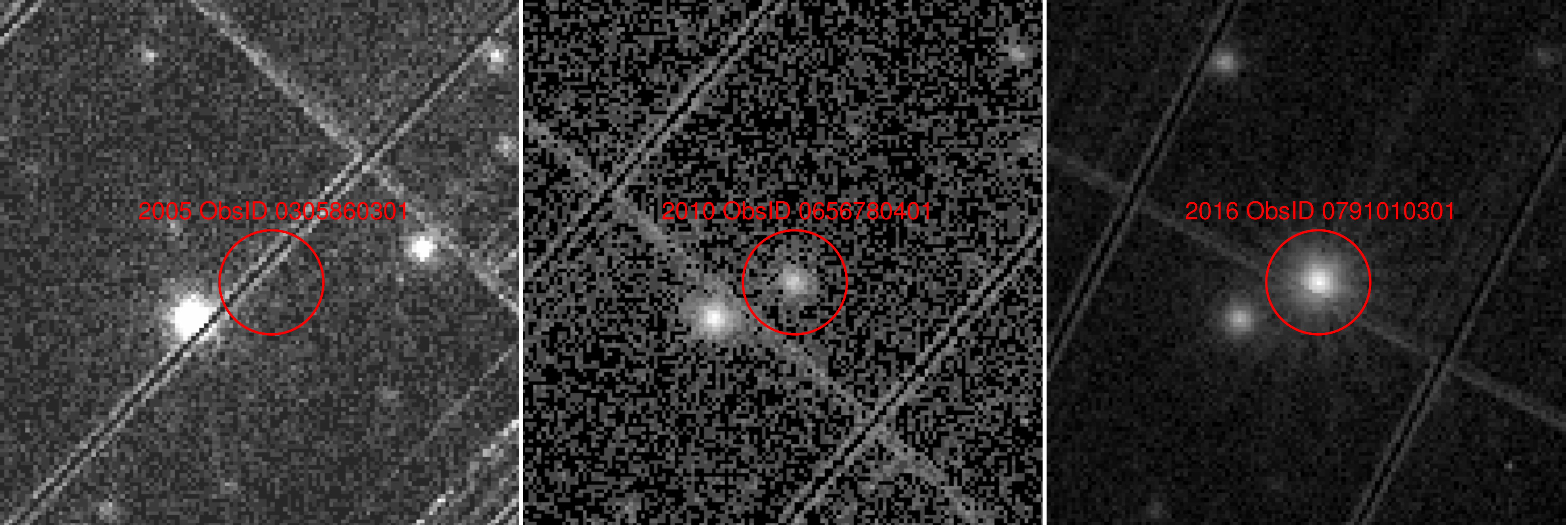} 
 \caption{\xmm\ EPIC-pn images of \ngc\ at different epochs marked in the figure. The red circles denote the source extraction region used for the analysis.}
   \label{fig1}
\end{center}
\end{figure}

\ngc\ is the fourth to be discovered in this class, and is the newly identified Ultra-luminous X-ray pulsar (ULXP) in NGC 300, located at a distance of 1.88 Mpc (\cite[Carpano \etal\ 2018b]{carpano2018}). 
The system was initially discovered in the optical wavelengths as a supernova in 2010 (\cite[Monard 2010]{monard2010}), but was classified as a supernova impostor event due to the high X-ray flux associated with a brightening in the optical and infrared regime (\cite[Binder \etal\ 2011]{binder2011}). \ngc\  was later identified as a possible supergiant B[e] HMXB owing to the spectroscopic and photometric information in the UV
and infrared wavelengths (\cite[Lau \etal\ 2016]{lau2016}).  \ngc\ was observed serendipitously in December 2016  in two consecutive \xmm\ observations (Obsid 0791010101 and 0791010301) performed simultaneously with \nus\ (Obsid 90401005002). The \xmm\  observations were performed for a duration of 139+82 ks and the \nus\  observation had an exposure of 163\,ks. \ngc\ was detected in its brightest state ever during this simultaneous observing campaign.
 \ngc\  was also observed with \xmm\ in 2010 when the source was reported in outburst for the first time (Obsid 0656780401), and in 4 other observations from 2000 to 2005 when the source was in the field of view but was not detected. Fig.\,\ref{fig1} shows the \xmm\ EPIC-pn image of \ngc\ during the 2005, 2010 (Obsid  0305860301) and 2016 observations.

\section{Discovery of pulsations and the extreme spin-period evolution of \ngc\ }
\begin{figure}
\centering
  \resizebox{0.52\hsize}{!}{\includegraphics[angle=-90,clip=]{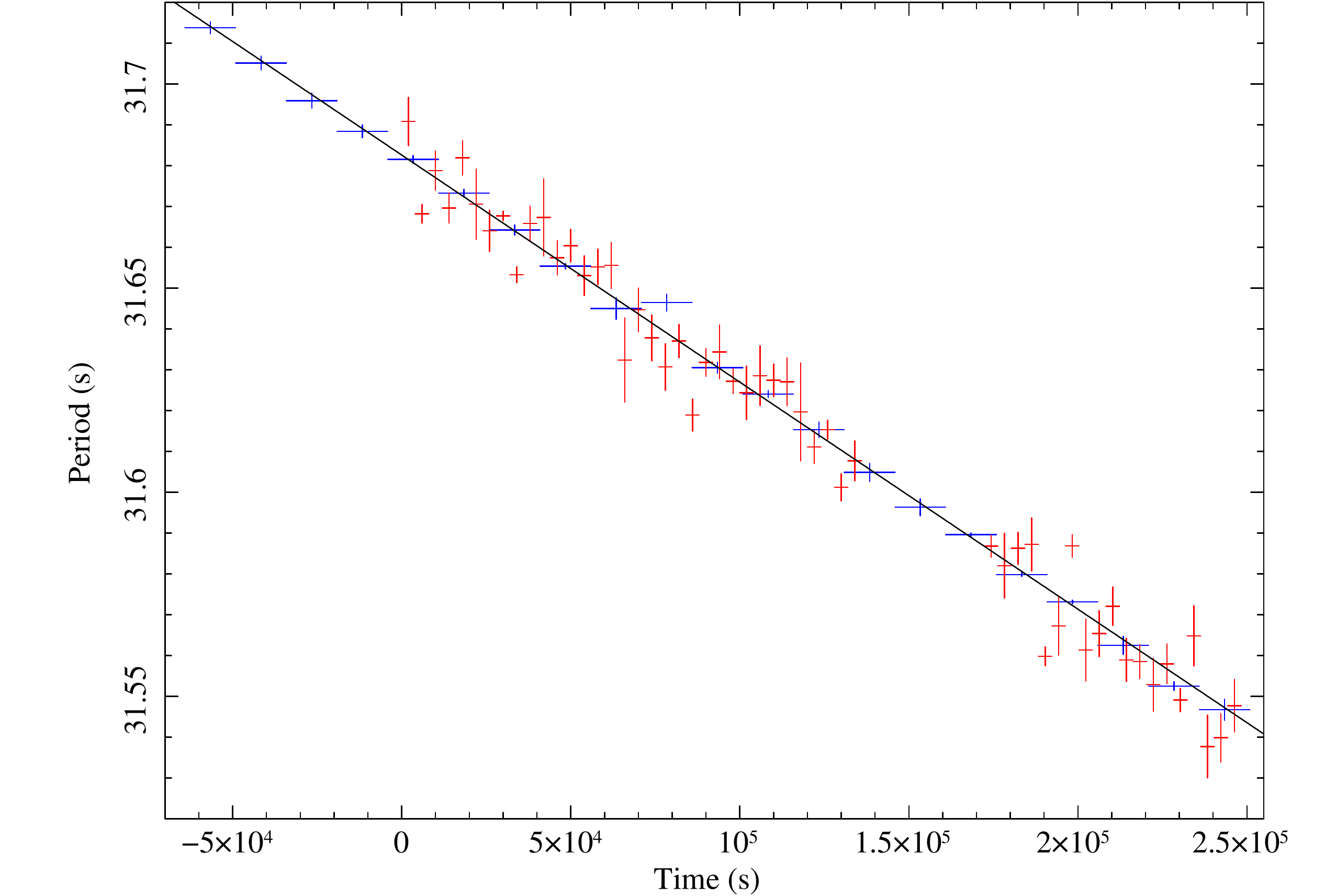}}
  
  \caption{
    Spin period evolution of \ngc\ obtained from 4\,ks intervals of EPIC-pn (red crosses) and 15\,ks intervals 
    of \nus\ data (blue crosses). The straight line represents the best-fit model of a linear period decrease applied 
    to both data sets. Time zero corresponds to the start of the EPIC-pn exposure. Figure is taken from \cite[Carpano \etal\ 2018b]{carpano2018}.
  }
  \label{fig:pulse}
\end{figure}

 Using the 2016 observations, \cite[Carpano \etal\ 2018a]{carpano2018a}  reported the discovery of a strong periodic modulation in the X-ray flux with a pulse period of 31.6 s and a very rapid spin-up rate.
 A refined timing analysis was further performed using a Bayesian method (see \cite[Carpano \etal\ 2018b]{carpano2018}), to probe the spin period evolution in detail. The  EPIC-pn data was split into 4\,ks intervals (53 intervals from the two observations,  0.2$-$10\,keV band), and the \nus\ data was split into in 21 intervals of 15\,ks in an energy band of 3$-$20 keV. The evolution of the spin period is shown in Fig.~\ref{fig:pulse}. The spin period of \ngc\ decreased linearly from 
$\sim$31.71\,s at the start of the \nus\ observation to $\sim$31.54\,s at the end of the \xmm/\nus\ observations.
The period derivative inferred from a model with a constant and linear term fitted to the \xmm\ and \nus\ data is 
($-$5.563$\pm$0.024)\expo{-7}\,s s$^{-1}$ with a spin period of 31.68262$\pm$0.00036\,s 
at the start of the of the EPIC-pn exposure (MJD 57739.39755). 
The pulsed fraction (0.2$-$10\,keV),  which was defined as the proportion of 
flux integrated over the pulse profile above minimum flux relative to the total integrated flux, 
increased slightly from 56.3$\pm$0.3\% during the first 2016 \xmm\ observation to 57.4$\pm$0.3\% in the second. 
The pulsed fraction also increased strongly with energy with 72.1$\pm$0.4\% in the \nus\ data (3$-$20\,keV).

\begin{figure}
\centering
    \resizebox{0.5\hsize}{!}{\includegraphics[angle=-90,clip=]{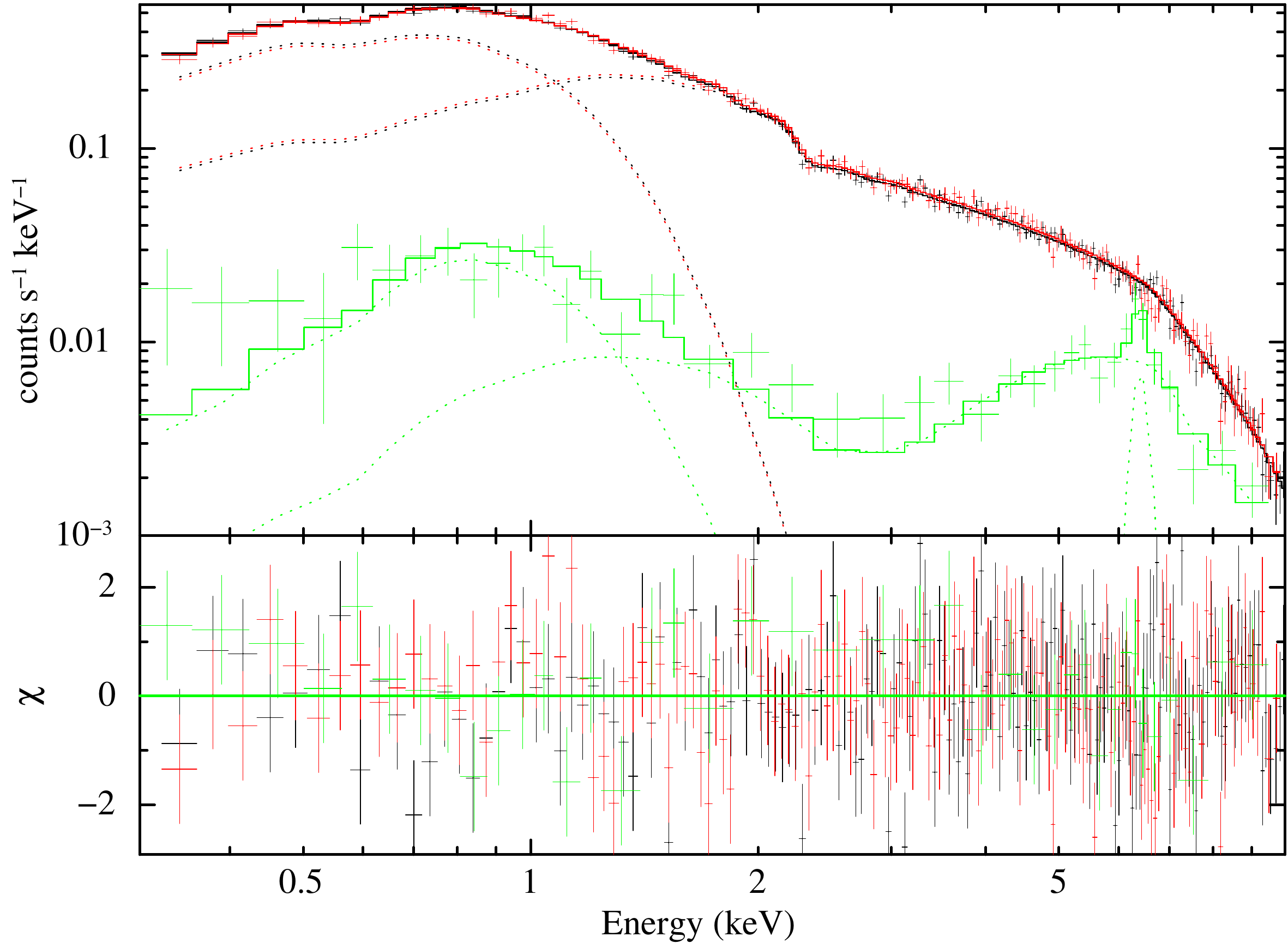}}
  \caption{
   Simultaneous spectral fit of \ngc\ using the EPIC-pn spectra together with the residuals as above. 
   Observation 0791010101 is marked in black, 0791010301 in red and 0656780401 (from 2010) in green. Figure is taken from \cite[Carpano \etal\ 2018b]{carpano2018}.
  }
  \label{fig:pulse1}
\end{figure}
\section{Comparison between the 2010 and 2016  \xmm\ X-ray spectra}
  The broadband X-ray spectra were fit  with a two-component model consisting of a power law with high-energy cutoff and a soft thermal 
component (disk black-body). The details are provided in \cite[Carpano \etal\ 2018b]{carpano2018}. The residuals after fitting this model indicated the presence of a further softer spectral component which can be attributed to the scattering 
and reprocessing of the X-ray photons originating in the vicinity of the neutron star, by an additional absorbing component. 
This was modelled using a partial-covering absorber component applied to the power-law and black-body 
components together. The model represents a physical scenario where the underlying continuum consists of a combination of power-law component 
(originating from the vicinity of the neutron star) plus a  disk black-body component (originating from the inner accretion disk), 
modified by scattering and absorption by additional material (most likely located in the clumpy wind of the supergiant companion or 
inner part of the circum-stellar disk of a Be star). The broadband unabsorbed luminosity of the source in 2016 (\xmm\ + \nus\ observations) was 4.7$\times10^{39}$ \ergs\, in the 
energy range of 0.3--30\,keV. 

The \xmm\ spectrum taken in 2010 was drastically different compared to that in 2016, with a soft component seen at energies $<$2 keV,
an almost flat spectrum between 2--4 keV and a bump-like feature above 5\,keV. This can be explained if the column density 
was significantly higher in 2010 and the direct component of the emission was reduced drastically. We investigated this by performing a simultaneous
spectral fit of the three EPIC-pn observations, assuming the same
underlying continuum spectrum as used in the broad-band spectral fit, and allowing all the absorption components to vary. In order to account for an intrinsic variation in the X-ray luminosity of the source, 
the power-law normalisation was also left to vary. The spectra and the best-fit model are shown in Fig.~\ref{fig:pulse1}. The 2010 \xmm\  observation can be explained by a similar intrinsic luminosity, but affected by a high partial absorption (i.e. equivalent hydrogen column density  $\ensuremath{N_{\mathrm{H}}} \sim 10^{23}$ cm$^{-2}$), compared to the 2016 observations where the partial absorption component was lower by a factor of 100.
\section{Spin and count rate evolution from the \swi\ and {\it NICER} observations}
\begin{figure}
\centering
  \resizebox{0.79\hsize}{!}{\includegraphics[angle=0,clip=]{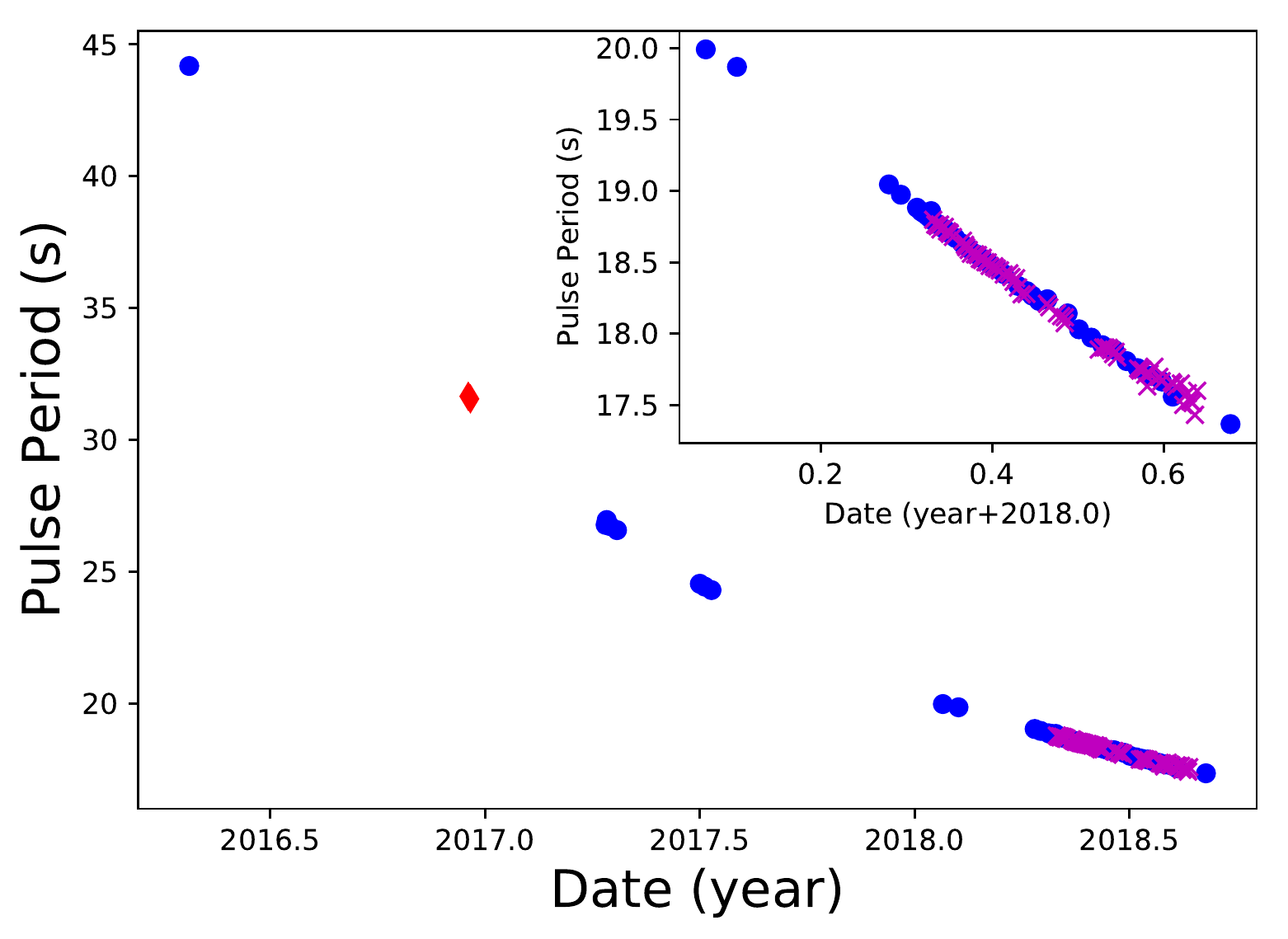}}
  
  \caption{Long-term time evolution of the pulse period of
\ngc\ from 2016-04-25 to 2018-08-22. The \swi\ observations are marked in blue, and the {\it NICER} observations in magenta.
The simultaneous \xmm\ /\nus\, observation is marked in red.}
  \label{test}
\end{figure}

\begin{figure}
\centering
  \resizebox{0.79\hsize}{!}{\includegraphics[angle=-90,clip=]{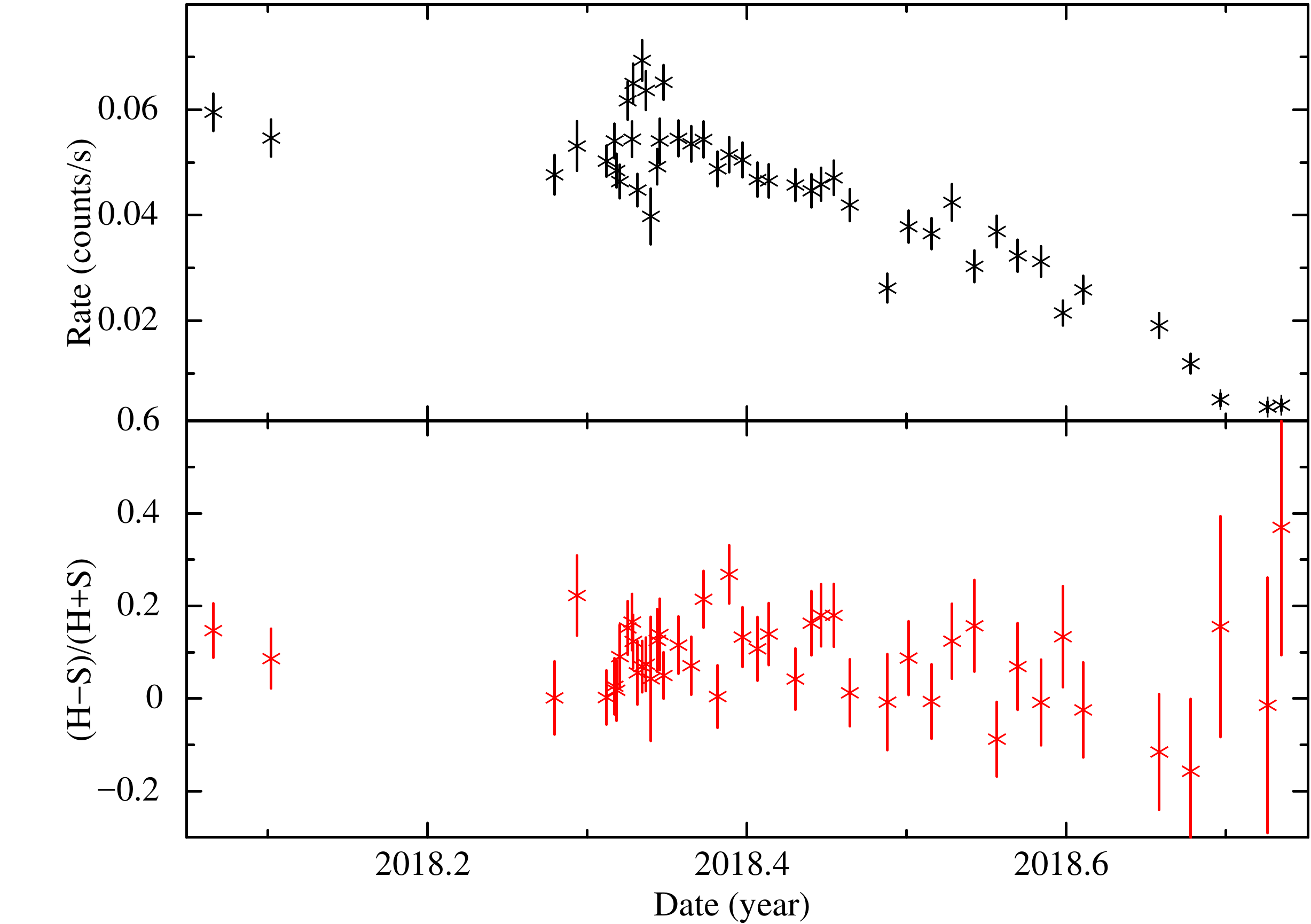}}
  
  \caption{Evolution of the count rate (0.2--10\,keV) and the hardness ratio using from the \swi\  observations corresponding to the inset in Fig.~\ref{test} (2018-01-24 to 2018-08-22).
  The hardness ratio is defined as shown in the second panel of the figure. The soft band (S) corresponds to 0.2--1.5\,keV and the hard band (H) to 1.5--10\,keV.}
  \label{swift}
\end{figure}

Given the extreme spin-up rate of \ngc\ as derived from \cite[Carpano \etal\ 2018b]{carpano2018}, the system is a rare opportunity to probe the relation between the accretion torque and the spin-up of a neutron star at super-critical mass accretion rates. While a detailed study presenting the spin evolution of \ngc\, and comparison with standard accretion torque models will be presented in a forthcoming paper (\cite[Vasilopoulos \etal\ 2018]{vas2018}), we present here the spin-up history of the source using data from the \swi /XRT and {\it NICER} monitoring campaigns of \ngc\ . The data spans from MJD 57502.275375 to MJD 58352.427225 where the source spins up from $\sim$45 to $\sim$17.5 s.  Fig.~\ref{test} shows the spin evolution of \ngc\ indicating a steady spin-up of the source for a span of 2.3 years. Fig.~\ref{swift} shows the recent count rate and hardness ratio evolution of the source starting from 2018. Although the count rate exhibits a gradual decline, \ngc\ continues to spin up.
\section{Summary}
\ngc\ was discovered as the fourth ULXP, exhibiting an extreme spin-up rate and a relatively constant luminosity over a long span of time. 
The secular spin period derivative of $-$5.56\expo{-7}\,s s$^{-1}$ seen over three days 
is one of the highest ever observed from an accreting neutron star, and the strong spin evolution is further supported by archival \swi /XRT and {\it NICER} observations
spanning over 2 years. The broadband X-ray spectrum derived from the 2016 observations is similar to what is observed from supergiant HMXBs, although the power law is quite steep and the high-energy 
cutoff starts at a relatively low energy. The archival \xmm\ spectrum from 2010 can be modelled by the same continuum, however requires a much higher absorption.
This indicates that \ngc\ was also in the ULX state at similar intrinsic X-ray luminosity, albeit highly absorbed in 2010.  \ngc\ provides a rare opportunity to probe the spin evolution of an accreting neutron star at extreme
accretion rates, and to understand the similarities between ULXPs and supergiant HMXBs.

\end{document}